# Propagation of Electromagnetic Waves in Neutron Star

Sania Saleem[1], Saeeda Sajjad[1], Samina Masood[2]
[1]*Department of Space Science, Institute of Space Technology, Pakistan*
[2]*Department of Physical and Applied Sciences, University of Houston- Clear Lake, TX, USA*

*Abstract—* **Propagation properties of electromagnetic (EM) waves in the dense medium of neutron star are studied. It is pointed out that EM waves develop a longitudinal component when they propagate in such a dense medium. Renormalization scheme of quantum electrodynamics (QED) is used to investigate the behavior of EM waves in transverse and longitudinal directions. Medium response to EM waves indicates that the electromagnetic properties of the dense material are modified as a result of the interaction of EM waves with the matter. Using QED, expressions for the electromagnetic properties such as electric permittivity, magnetic permeability and the refractive index of dense matter are obtained. The results are applied to a particular star for illustration.**

*Keywords: Chemical potential, Electromagnetic wave, Neutron star, Propagation.*

## I. INTRODUCTION

Neutron stars are unique example of extremely dense objects in the universe and they contain most of the matter in the form of neutrons. Material in these stars is in the highly degenerate form and density is comparable to atomic nuclei. Fermi temperature in these stars is very high and particles have ultra-relativistic energies. The quantum interactions along with ultra-relativistic particle dynamics make such a material very complex. Pulsars are also believed to be rotating neutron stars which have very large magnetic fields of the order of $10^{12}$G and emit radiation in a wide spectrum ranging from radio waves to x-rays. Statistical properties of the material in these objects are completely different from ordinary matter encountered on Earth. Neutron stars may have small percentage of electrons and protons which have not been combined to produce neutrons due to lack of energy [1, 2].

Aim of this work is to understand the propagation characteristics of electromagnetic waves and response of the medium in which electrons are expected to have large chemical potential. For this purpose renormalization scheme of quantum electrodynamics (QED) is invoked

Perturbation techniques of QED are employed to calculate the vacuum polarization tensor of QED. Medium properties are modified significantly due to the interaction of longitudinal component of the electromagnetic signals with hot and dense matter of neutron star having large chemical potential. The electromagnetic properties of the medium turn out to be a function of temperature (T) and chemical potential of electrons ($\mu_e$).

In 2017 [3] a theoretical model based on QED was presented to evaluate polarization tensor of photons in a hypothetical system in which electron temperature was assumed to be much larger than the background. Matter in such a state is believed to exist in super dense stars.

This theoretical model is useful to investigate the behavior of subatomic particles as a source for magnetic fields. Electromagnetic (EM) characteristics of the stellar medium are investigated in terms of electric permittivity (ϵ), magnetic permeability (μ) and refractive index (n) of the medium. We use some of the already existing results of QED in an extremely hot and dense medium and estimate the important physical parameters numerically.

## II. CALCULATION SCHEME

Here we present the theoretical results [4] based on QED renormalization and apply the results to neutron star medium at high temperature and density. The vacuum propagator is replaced with the hot and dense propagators modified by the Bose-Einstein distribution for bosons and Fermi-Dirac distribution for fermions. Following this procedure, the most general form of the vacuum polarization tensor can be written as [5, 6]

$$\pi_{\mu\nu}(K,\mu_e) = \iota e^2 \int \frac{d^4 p}{(4\pi)^4} T_r\{\gamma_\mu(\gamma_\alpha p^\alpha + \gamma_\alpha K^\alpha + m)\gamma_\nu(\gamma_\alpha p^\alpha + m)\} \left[\frac{1}{(p+K)^2 - m^2} + \Gamma_F(p+K,\mu_e)\right] \left[\frac{1}{p^2-m^2} + \Gamma_F(p,\mu_e)\right],$$

$$\pi_{\mu\nu}(K,\mu_e) = \iota e^2 \int \frac{d^4 p}{(4\pi)^4} T_r\{\gamma_\mu(\gamma_\alpha p^\alpha + \gamma_\alpha K^\alpha + m)\gamma_\nu(\gamma_\alpha p^\alpha + m)\} \left[\frac{1}{(p+K)^2 - m^2} + \Gamma_F(p+K,\mu_e)\right] \left[\frac{1}{p^2-m^2} + \Gamma_F(p,\mu_e)\right], \quad (1)$$

Here [7],
$$\Gamma_F(p,\mu_e) = 2\pi\iota\delta(p^2 - m^2)[\theta(p_0)n_f(p,\mu_e) + \theta(-p_0)n_f(p,-\mu_e)] \quad (2)$$

Here, K the 4-Momentum of photon is satisfying the expressions,
$K^2 = \omega^2 - k^2$.
where
$$\omega = K_\alpha u^\alpha \quad \text{and,} \quad u^\alpha = (1,0,0,0) \quad (3)$$

The polarization tensor has two parts,
$$\pi_{\mu\nu}(K) = \pi_{\mu\nu}^{T=0}(K) + \pi_{\mu\nu}^{\beta}(K, \mu_e). \quad (4)$$
where $\pi_{\mu\nu}^{\beta}(K, \mu_e)$ denotes the medium contribution and the other tensor $\pi_{\mu\nu}^{T=0}(K)$ denotes vacuum contribution. The medium contribution is given as follows,

$$\pi_{\mu\nu}^{\beta}(K, \mu_e) = -\frac{2\pi e^2}{2}\int \frac{d^4p}{(4\pi)^4} T_\tau \{\gamma_\mu(\gamma_\alpha p^\alpha + \gamma_\alpha K^\alpha + m)\gamma_\mu(\gamma_\alpha p^\alpha + m)\}[\frac{\delta[(p+K)^2 - m^2]}{p^2 - m^2}\{n_f(p+K,\mu_e) + n_f(p+K,-\mu_e)\} + \frac{\delta[(p-m)^2]}{(p+K)^2 - m^2}\{n_f(p,\mu_e) + n_f(p,-\mu_e)\}], \quad (5)$$

The medium contribution term $\pi_{\mu\nu}^{\beta}(K, \mu_e)$ contains both transverse component $\pi_T(k, \omega)$ and longitudinal component $\pi_L(k, \omega)$ as,,

$$\pi_{\mu\nu}(K, \mu_e) = P_{\mu\nu}\pi_T(K, \mu_e) + Q_{\mu\nu}\pi_L(K, \mu_e). \quad (6)$$
Where the transverse part of the vacuum polarization tensor is,
$$P_{\mu\nu} = \widetilde{g}_{\mu\nu} + \frac{\widetilde{K}_\mu \widetilde{K}_\nu}{k^2}. \quad (7a)$$
And the longitudinal component is given as,
$$Q_{\mu\nu} = -\frac{1}{K^2 k^2}(k^2 u_\mu + \omega \widetilde{K}_\mu)(k^2 u_\nu + \omega \widetilde{K}_\nu). \quad (7b)$$
Also,
$$\widetilde{g}_{\mu\nu} = g_{\mu\nu} - u_\mu u_\nu \quad (7c)$$
Here,
$$\widetilde{K}_\mu = K_\mu - \omega u_\mu \quad (7d)$$
So that they meet the conditions of,

$$P_\nu^\mu P_\alpha^\mu = P_\alpha^\mu. \quad (8a)$$
$$Q_\alpha^\mu = Q_\nu^\mu Q_\alpha^\nu. \quad (8b)$$
$$K_\mu P_\nu^\mu = 0. \quad (8c)$$
$$K_\mu Q_\nu^\mu = 0. \quad (8d)$$

We use the well-known calculated results of longitudinal and transverse components of the vacuum polarization tensors for different ranges of chemical potential and temperature. We have the super dense medium of neutron star so we use the limit of $(T \ll \mu_e)$, Where, T is the temperature and $\mu_e$ is the chemical potential.

In the limit $(T \ll \mu_e)$, we have
$$\pi_L(K, T, \mu_e) \cong \frac{e^2}{2\pi^2 k}(1 - \frac{\omega^2}{k^2})[(k - 2\omega \ln\frac{\omega-k}{\omega+k})J_1 - 2A_1 J_2 - \frac{1}{2}A_2 J_3. \quad (9a)$$

And,
$$\pi_T(K, T, \mu_e) \cong \frac{e^2}{2\pi^2 k}[\{\frac{\omega^2}{k} + \omega(1 - \frac{\omega^2}{k^2})\ln\frac{\omega+k}{\omega-k}\}J_1 + \{\frac{kK^2 + km^2}{2} + A_1(1 - \frac{\omega^2}{k^2})\}J_2 + \{\frac{k^3}{12} + \frac{K^2 k}{24m^2}(3\omega^2 + k^2)\} + (1 - \frac{\omega^2}{k^2})\frac{A_2}{4m^2} - \frac{m^2(5\omega^2 - 3k^2)}{8k}J_0] \quad (9b)$$

where $\pi_L(K, T, \mu_e)$ represents the longitudinal component and $\pi_T(K, T, \mu_e)$ represents the transverse component of the polarization tensors. Expressions for $A_1$, $A_2$, $J_2$ and $J_3$ are as follows,

$$A_1 = \frac{k^3}{12} + \frac{\omega^2 k}{2} + \frac{m^2 k}{K^2}(\omega^2 + k^2) + \frac{kK^2}{2} - \omega^2 k[\frac{2m^2}{k^2} - 1]. \quad (10a)$$

$$A_2 = \frac{1}{4}[\frac{m^2}{K^4}\{4m^2 k^3(4\omega^2 + k^2) - K^2(3k\omega^4 + 12\omega^2 k^3 + k^5)\} + k[\omega^4 + 20\omega^2 k^2 + 2k^4 + \frac{m^4}{2}] + \frac{K^2 k]}{6}(3\omega^2 + k^2)] \quad (10b)$$

$$J_1 = \frac{1}{2}[\frac{\mu_e^2}{2}[1 - \frac{m^2}{\mu_e^2}] + \frac{1}{\beta}\{\alpha(m\beta, \mu_e) - \alpha`(m\beta, \mu_e)\} - \frac{1}{\beta^2}\{c(m\beta, \mu_e) + c`(m\beta, \mu_e)\}]. \quad (11a)$$

$$J_2 = \frac{1}{2}[\ln\frac{\mu_e}{m} + b(m\beta, \mu_e) - b`(m\beta, \mu_e)]. \quad (11b)$$

$$J_3 = \frac{1}{2}[\frac{1}{2m^2}[1 - \frac{m^2}{\mu_e^2}] - \frac{1}{4\mu_e^2} + \frac{1}{m^2}\{n_f(\mu_e + m) + n_f(\mu_e - m)\} + \frac{\beta}{m}\{\frac{e^{-\beta(\mu_e + m)}}{[1 + e^{\beta(\mu_e + m)}]^2} + \frac{e^{-\beta(\mu_e - m)}}{[1 + e^{-\beta(\mu_e - m)}]^2}\} + d(m\beta, \mu_e) + d`(m\beta, \mu_e)]. \quad (11c)$$

The nonzero terms in the longitudinal component show that the waves acquire dynamically generated mass which modifies the electromagnetic properties of the dense medium. We analyze these calculations in the limit ($\omega \gg k$), where $K^2 = \omega^2$, the other parameters are given as follows,

$$A_1 = \frac{-2m^2\omega^2}{k} \quad \text{and} \quad A_2 = \frac{3\omega^4 k}{8} \quad (12),$$
$$j_1 = \frac{1}{2}[\frac{\mu_e^2}{2} - \frac{m^2}{2}]. \quad (13a)$$
$$j_2 = \frac{1}{2}\ln\frac{\mu_e}{m}. \quad (13b)$$
$$j_3 = \frac{1}{4m^2}. \quad (13c)$$

By inserting the values of $j_1$, $j_2$ and $j_3$, the electric permittivity of the medium turns out to be,

$$\varepsilon_E(k, \mu_e, T) = 1 - \frac{2\alpha}{\pi k K^2}[[k - 2\omega \ln\frac{\omega-k}{\omega+k}]j_1 - 2A_1 j_2 + \frac{1}{2}A_2 j_3][1 - \frac{\omega^2}{k^2}]. \quad (14a)$$

The magnetic permeability becomes,
$$\frac{1}{\mu_B(k, \mu_e, T)} \cong 1 + \frac{2\alpha}{\pi k^3 K^2}[[\frac{k^2}{2} + \omega^2][1 - \frac{\omega^2}{k^2}]\{2\omega \ln\frac{\omega-k}{\omega+k}j_1 - 2A_1 j_2 - \frac{1}{2}A_2 j_3\} - \frac{k^3 K^2 + k^4 m}{4}j_2 - \frac{m^2}{2}\{\frac{k^5}{12} + \frac{(3\omega^2 + k^2)k^3 K^2}{24m^2} + \frac{m^2 k^2(5\omega^2 - 3k^2)}{8k}\}j_3. \quad (14b)$$

The expression for the velocity of propagation of electromagnetic waves in medium is written as,
$$v_{prop} = \sqrt{\frac{1}{\varepsilon_{(k)}\mu_{(k)}}}. \quad (15)$$

In vacuum, Eq. (15) reduces to,

$$v_{prop} = c = \sqrt{\frac{1}{\varepsilon_{(0)}\mu_{(0)}}}. \quad (16)$$

Here $\varepsilon_{(0)}, \mu_{(0)}$ are values of electric permittivity and magnetic permeability in vacuum.

The index of refraction of the medium is calculated using,
$$n = \frac{c}{v_{prop}}. \quad (17)$$

Equation (16) and (17) yields,

$$n = \sqrt{\frac{\varepsilon_{(k)}\mu_{(k)}}{\varepsilon_{(0)}\mu_{(0)}}}. \quad (18)$$

Or,

$$n = \sqrt{\varepsilon_R \mu_R}. \quad (19)$$

The modifications in the parameters of electromagnetic properties of the medium of neutron star are analyzed numerically in the next section.

## III. NUMERICAL CALCULATIONS

We have investigated the dependence of electric permittivity ($\varepsilon_R$), magnetic permeability ($\mu_B$) and refractive index (n) on the energy and momentum of the photon for large values of electron chemical potential ($\mu_e$) in neutron star for the case $\omega \gg k$. We cannot predict the exact value of chemical potential at any particular value of $\omega$ because the exact measurement of these parameters in such a dense medium is not possible. The current existing models and observational data lead to the maximum value of chemical potential to be around 250MeV in neutron star [2]. The variation in electric permittivity and magnetic permeability as a function of $\mu_e$, $\omega$ and k, indicate the distribution of matter in star and hence using the present approach, understanding of the structure of neutron star can be improved. All of these parameters are given in units of electron mass $m_e$ = 0.51 MeV. We have plotted graphs corresponding to large values of $\mu_e$ from (100-250) MeV.

The Fig. (1) shows the relative permittivity as a function of omega $\omega$ (the energy of the photon), for fixed k = 0.5 $m_e$ (~0.25 MeV). Electric permittivity does not seem to change much with respect to the chemical potential but increases significantly with increasing values of omega. In particular, the permittivity of the medium tends to increase as the energy of photons increases beyond 20 MeV that is for $\omega$ > 20 MeV. For lower energies of photon, permittivity increases slowly with $\omega$. The dependence of relative permittivity on omega remains almost the same for different values of chemical potential.
.

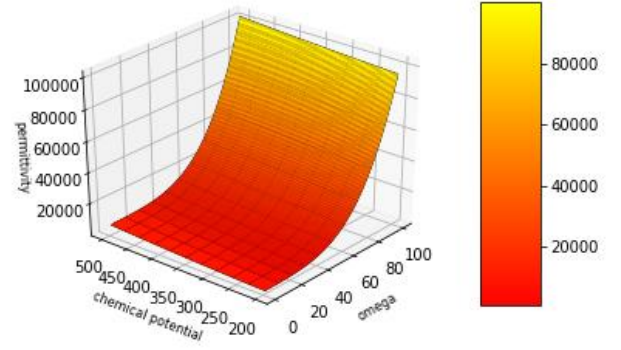

*Figure 1: Relative permittivity $\epsilon_R$ is plotted versus omega $\omega$ in units of electron mass $m_e$ for constant k*

The relative permeability on the other hand has opposite trend as shown in Fig. (2) as compared to Fig. This figure gives a plot of relative permeability as a function of omega at fixed k= (0.25 MeV) for different values of ($\mu_e$). It is obvious in the plot that relative permeability rapidly drops down at lower energy values and becomes constant beyond 1 MeV. The relative permeability has higher values for low energy photons and falls rapidly until the value reaches closer to a few MeV. The overall behavior of relative permeability is the same at all chemical potentials but at lower energies, the relative permeability is larger for low chemical potential (see for example $\mu_e$=100 MeV) in comparison with high chemical potential (for instance $\mu_e$=250 MeV)

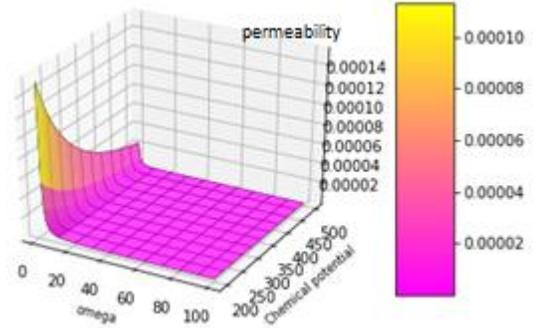

*Figure 2: Relative permeability $\mu_B$ is plotted versus omega $\omega$ in units of electron mass $m_e$ for constant k*

The relative permittivity has a different trend when plotted as a function of k (the momentum of the photon) at fixed omega = 10$m_e$ (5 MeV) as can be seen in Fig.3. We plot the relative permittivity for different values of k in the limit (k<<$\omega$). It shows that the permittivity is high at regions having smaller values of momentum and as the momentum increases the permittivity decreases and becomes almost constant beyond 0.5 MeV. The relative permittivity shows the dependence on chemical potential only at lower k values. Above 0.5 MeV the permittivity is almost constant for all values of ($\mu_e$).

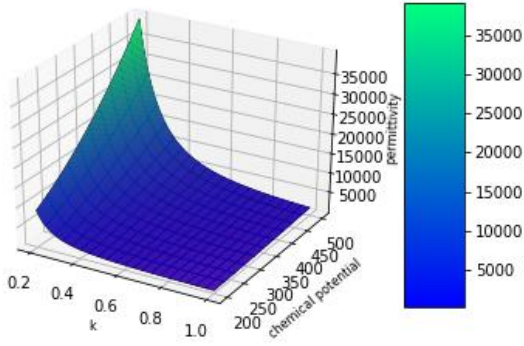

*Figure 3: Relative permittivity $\epsilon_R$ is plotted versus k in units of electron mass $m_e$ for constant $\omega$*

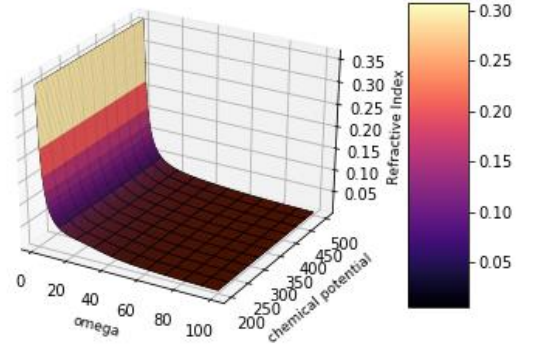

*Figure 5: Index of refraction is plotted versus omega ($\omega$) in units of electron mass $m_e$ for constant k*

The relative permeability is plotted in Fig.4 as a function of k (the momentum of photon) at constant omega = $10m_e$ (5 MeV). It shows that the magnetic permeability increases with the increase in photon momentum. The permeability is around zero at lower momenta of around 0.15 MeV but increases afterwards depending on the chemical potential. The relative permeability of the medium increases at lower values of chemical potential say about 100MeV

Refractive index shows a different behavior when plotted against the momentum (k) of photon at omega =10 $m_e$ ($\omega$ =5 MeV) for different values of chemical potentials as can be seen in Fig. 6. It is clear in this figure that the refractive index increases with momentum (k). Thus, it is clear that as momentum of photons increases, the light travels slowly.
Also the refractive index is independent of the chemical potential and increases for all values of $\mu_e$.

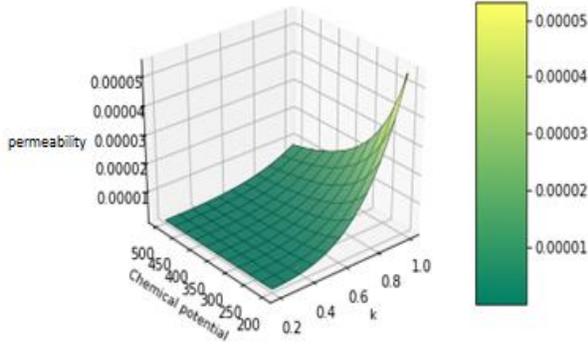

*Figure 4: Relative permeability $\mu_B$ is plotted versus k in units of electron mass $m_e$ for constant $\omega$*

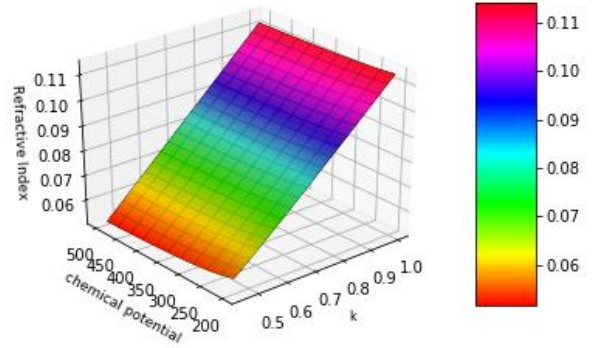

*Figure 6: Index of refraction is plotted versus k in units of electron mass $m_e$ for constant $\omega$*

We further explore the properties of the medium by studying the behavior of the refractive index n as a function of $\omega$, k and $\mu_e$. In Figure 5 we have plotted the refractive index as a function of omega (the energy of the photon) at k=0.25 MeV for different values of chemical potentials. It shows higher value of n corresponding to lower energy photons and decreases as energy of photons increases. It decreases steeply with increasing $\omega$ and at about 10MeV it becomes constant.
.

## IV.  SUMMARY

The propagation of electromagnetic waves in the dense matter of neutron star has been studied with the help of the renormalization scheme of quantum electrodynamics (QED).It has been pointed out that under the extreme conditions of macroscopic quantities; temperature, density and magnetic field in neutron stars,  the electromagnetic properties of the matter become function of the chemical potential $\mu_e$. We have found that the relative permittivity increases with the increase in photon energy ($\omega$). On the other hand the relative permittivity decreases with increase in photon momentum (k). The relative permeability seems to vanish as the photon energy increases beyond 2.5MeV. On the hand, relative permeability tends to increase with (k).

V.  REFERENCES

[1]  B. W. Carroll and D. A. Ostlie, *An introduction to modern astrophysics*: Cambridge University Press, 2017.

[2]  S. L. Shapiro and S. A. Teukolsky, *Black holes, white dwarfs, and neutron stars: The physics of compact objects*: John Wiley & Sons, 2008.

[3]  S. Masood and I. Saleem, "Propagation of electromagnetic waves in extremely dense media," *International Journal of Modern Physics A,* vol. 32, p. 1750081, 2017.

[4]  K. Ahmed and S. S. Masood, "Vacuum polarization at finite temperature and density in QED," *Annals of physics,* vol. 207, pp. 460-473, 1991.

[5]  S. S. Masood, "Propagation of monochromatic light in a hot and dense medium," *The European Physical Journal C,* vol. 77, p. 826, 2017.

[6]  S. S. Masood, "Renormalization of QED in superdense media," *Physical Review D,* vol. 47, p. 648, 1993.

[7]  E. J. Levinson and D. H. Boal, "Self-energy corrections to fermions in the presence of a thermal background," *Physical Review D,* vol. 31, p. 3280, 1985.